# Talking Condition Identification Using Second-Order Hidden Markov Models

Ismail Shahin

Electrical and Computer Engineering Department

University of Sharjah

ismail@sharjah.ac.ae

*Abstract*—**This work focuses on enhancing the performance of text-dependent and speaker-dependent talking condition identification systems using second-order hidden Markov models (HMM2s). Our results show that the talking condition identification performance based on HMM2s has been improved significantly compared to first-order hidden Markov models (HMM1s). Our talking conditions in this work are neutral, shouted, loud, angry, happy, and fear.**

## I. INTRODUCTION

Talking condition identification is the process of determining from which of the registered talking conditions an unknown talking condition comes.

Talking condition identification systems typically operate in one of two cases: text-dependent case or text-independent case. In the text-dependent case, utterances of the same text are used for both training and testing. In the text-independent case, training and testing involve utterances from different texts. The process of talking condition identification can be divided into two categories: "open set" and "closed set". In the "open set" category, a reference model for an unknown talking condition may not exist; however, in the "closed set" category, a reference model for an unknown talking condition is available.

Hidden Markov models (HMMs) are considered in the last three decades as one of the most widely used modeling techniques in the fields of speech recognition and speaker recognition [1]. HMMs use Markov chain to model the changing statistical characteristics that exist in the actual observations of speech signals. The Markov process is a double stochastic process where there is an unobservable Markov chain defined by a state transition matrix. Each state of the Markov chain is associated with either a discrete output probability distribution (discrete HMMs) or a continuous output probability density function (continuous HMMs) [2].

HMMs are powerful models in optimizing the parameters that are used in modeling speech signals. This optimization decreases the computational complexity in the decoding procedure and improves the recognition accuracy [2].

## II. MOTIVATION

A major motivation of talking condition identification systems comes from the desire to develop human machine interface that is more adaptive and responsive to a user's identity. The main task of intelligent human-machine interaction is to empower a computer with the affective computing ability so that a computer can recognize speakers for many different applications.

Most of the people who work in the areas of speech recognition and speaker recognition focus their work on speech under the neutral talking condition and few people focus their work on speech under stressful and emotional talking conditions. Stressful and emotional talking conditions can be defined as talking conditions that cause speakers to vary their production of speech from the neutral talking condition. The neutral talking condition is defined as the talking condition in which speech is produced assuming that the speakers uttering their voice in a "quiet room" with no task obligations. Some talking conditions are designed to simulate speech produced by different speakers under real stressful and emotional talking conditions.

There were some work on emotion or talking condition recognition using HMM1s. Li and Zhao used three classification methods to recognize emotions in speech. These methods include vector quantization, artificial neural networks, and Gaussian mixture density model [3]. Zhao *et al.* used HMM1s in studying the nonlinear feature based classification of speech under stress [4]. DeSilva and Ng described the use of statistical techniques and HMM1s in the recognition of emotions [5]. Nogueiras *et al.* used semi-continuous HMM1s in emotion recognition [6]. Shahin used first-order circular hidden Markov models to enhance talking condition identification performance [7].

Our work in this paper focuses on identifying an unknown talking condition of text-dependent and speaker-dependent talking condition identification systems based on HMM2s.

The remainder of this paper is structured as follows. Section III explains the details of second order hidden Markov models. Extended Viterbi and Baum-Welch algorithms are covered in Section IV. The speech database used in this work is explained in Section V. Results and discussion of our work are given in Section VI. Concluding remarks of this work appear in Section VII.

### III. SECOND ORDER HIDDEN MARKOV MODELS

New models called HMM2s were introduced and implemented under the neutral talking condition by Mari *et al.* [8].

In HMM1s, the underlying state sequence is a first-order Markov chain where the stochastic process is specified by a 2-D matrix of a priori transition probabilities ($a_{ij}$) between states $s_i$ and $s_j$ where $a_{ij}$ are given as,

$$a_{ij} = \text{Prob}\left(q_t = s_j | q_{t-1} = s_i\right) \quad (1)$$

where $q_t$ denotes any state at time $t$.

In HMM2s, the underlying state sequence is a second-order Markov chain where the stochastic process is specified by a 3-D matrix ($a_{ijk}$). Therefore, the transition probabilities in HMM2s are given as [8],

$$a_{ijk} = \text{Prob}\left(q_t = s_k | q_{t-1} = s_j, q_{t-2} = s_i\right) \quad (2)$$

with the constraints,

$$\sum_{k=1}^{N} a_{ijk} = 1 \qquad N \geq i, j \geq 1$$

where $N$ is the number of states.

The probability of the state sequence, $Q \triangleq q_1, q_2, ..., q_T$, is defined as,

$$\text{Prob}(Q) = \Psi_{q_1} a_{q_1 q_2} \prod_{t=3}^{T} a_{q_{t-2} q_{t-1} q_t} \quad (3)$$

where $\Psi_i$ is the probability of a state $s_i$ at time $t = 1$, $a_{ij}$ is the probability of transition from a state $s_i$ to a state $s_j$ at time $t = 2$, and $T$ is the length or duration of an observation sequence, $O \triangleq O_1, O_2, ..., O_T$, and it is equal to the total number of frames.

Each state $s_i$ is associated with a mixture of Gaussian distributions,

$$b_i(O_t) \triangleq \sum_{m=1}^{M} c_{im} N(O_t; \mu_{im}, \Sigma_{im}), \text{ with } \sum_{m=1}^{M} c_{im} = 1 \quad (4)$$

where the vector $O_t$ is the input vector at time $t$.

Given a sequence of observed vectors, the joint state-output probability is defined as,

$$\text{Prob}(Q, O | \lambda) = \Psi_{q_1} b_{q_1}(O_1) a_{q_1 q_2} b_{q_2}(O_2).$$
$$\prod_{t=3}^{T} a_{q_{t-2} q_{t-1} q_t} b_{q_t}(O_t) \quad (5)$$

where $b_{q_t}(O_t)$ is the observation symbol probability given as,

$$b_{q_t}(O_t) = \text{Prob}(O_t \text{ emitted at } t | q_{t-1} = S_i)$$

## IV. EXTENDED VITERBI AND BAUM-WELCH ALGORITHM

The most likely state sequence can be found by using the probability of the partial alignment ending at a transition $(s_j, s_k)$ at times $(t-1, t)$,

$$\delta_t(j,k) \triangleq \text{Prob}\begin{pmatrix} q_1, \ldots, q_{t-1} = s_j, q_t = s_k, O_1, O_2, \ldots, \\ O_t | \lambda \end{pmatrix}$$
$$T \geq t \geq 2, N \geq j, k \geq 1 \quad (6)$$

Recursive computation is given by,

$$\delta_t(j,k) = \max_{N \geq i \geq 1} \{\delta_{t-1}(i,j) \cdot a_{ijk}\} \cdot b_k(O_t)$$
$$T \geq t \geq 3, N \geq j, k \geq 1 \quad (7)$$

The forward function $\alpha_t(j,k)$ defines the probability of the partial observation sequence, $O_1, O_2, \ldots, O_t$, and the transition $(s_j, s_k)$ between times $t-1$ and $t$ is given by,

$$\alpha_t(j,k) \triangleq \text{Prob}(O_1, \ldots, O_t, q_{t-1} = s_j, q_t = s_k | \lambda)$$
$$T \geq t \geq 2, N \geq j, k \geq 1 \quad (8)$$

$\alpha_t(j,k)$ can be computed from the two transitions: $(s_i, s_j)$ and $(s_j, s_k)$ between states $s_i$ and $s_k$ as,

$$\alpha_{t+1}(j,k) = \sum_{i=1}^{N} \alpha_t(i,j) \cdot a_{ijk} \cdot b_k(O_{t+1})$$
$$T - 1 \geq t \geq 2, N \geq j, k \geq 1 \quad (9)$$

The backward function $\beta_t(i,j)$ can be expressed as,

$$\beta_t(i,j) \triangleq \text{Prob}(O_{t+1}, \ldots, O_T | q_{t-1} = s_i, q_t = s_j, \lambda)$$
$$T - 1 \geq t \geq 2, N \geq i, j \geq 1 \quad (10)$$

where $\beta_t(i,j)$ is defined as the probability of the partial observation sequence from $t+1$ to $T$, given the model $\lambda$ and the transition $(s_i, s_j)$ between times $t-1$ and $t$.

## V. SPEECH DATABASE

Our speech database in this work was collected from 12 (6 males and 6 females) untrained healthy adult native speakers of American English. Each speaker uttered 8 sentences where each sentence was uttered 9 times (9 utterances or tokens) under each of the neutral, shouted, loud, angry, happy, and fear talking conditions. The 8 sentences were unbiased towards any talking condition. These sentences were:

1) He works five days a week.
2) The sun is shining.
3) The weather is fair.
4) The students study hard.
5) Assistant professors are looking for promotion.
6) University of Sharjah.
7) Electrical and Computer Engineering Department.
8) He has two sons and two daughters.

Our speech database was captured by a speech acquisition board using a 16-bit linear coding A/D converter and sampled at a sampling rate of 16 kHz. Our database was a 16-bit per sample linear data. Each sentence under each talking condition was then applied every 5 ms to a 30 ms Hamming window. 12*th* order linear prediction coefficients (LPCs) were extracted from each frame by the autocorrelation method. The 12*th* order LPCs were then transformed into 16*th* order linear prediction cepstral coefficients (LPCCs).

The LPCC feature analysis was used to form the observation vectors in each of HMM1s and HMM2s. The number of states, *N*, was 5. The number of mixture components, *M*, was 5 per state, with a continuous mixture observation density was selected for each of HMM1s and HMM2s. Our speech database in this work was a "closed set". Our database was divided into training data and test data.

In the training session, one reference model per speaker per sentence per talking condition was derived using 5 of the 9 utterances. Training of models in this session had been done separately based on each of HMM1s and HMM2s.

In the testing (identification) session, each one of the 12 speakers used 4 of the 9 utterances per the same speaker per the same sentence under each talking condition. This session had been done separately based on each of HMM1s and HMM2s. Based on the probability of generating an utterance, the model with the highest probability was chosen as the output of the talking condition identification system based on each of HMM1s and HMM2s.

VI. RESULTS AND DISCUSSION

Table I and Table III summarize the results of talking condition identification performance using our speech database based on HMM1s and HMM2s respectively.

Table II and Table IV show confusion matrices that represent the percentage of confusion of a test talking condition with the other talking conditions based on HMM1s and HMM2s respectively.

It is evident from Table I that the talking condition identification system identifies correctly the neutral talking condition with an average identification performance of 99%. Table I and Table II show that the most easily recognizable talking condition is the neutral talking condition (99%). On the other hand, the talking condition identification performance for the remaining talking conditions is significantly degraded based on HMM1s. This is because HMM1s are inefficient models under the stressful and emotional talking conditions since the changes in the statistical characteristics that exist in the actual observations of the stressful and emotional talking conditions are greater than that of the neutral talking condition.

Table I shows that the least talking condition identification performance occurs under the shouted talking condition. Hence, the least easily recognizable talking condition is the shouted talking condition (30%). The reason for this poor identification performance is that the shouted talking condition is considered as the most stressful talking condition [9]. It was reported in many publications that HMM1s did not perform well under such a talking condition [9].

Column 3 of Table II (for example) says that 34% of the utterances that were portrayed as shouted talking condition were evaluated as loud talking condition, 31% of the utterances that were portrayed as shouted talking condition were evaluated as angry talking condition, *etc…* Table V shows that HMM2s significantly enhance the shouted talking condition identification performance compared to that based on HMM1s. The improvement rate is 26.7%. Comparing Table II with Table IV, the shouted talking condition has the highest confusion percentage with each of the loud and angry talking conditions.

Using HMM1s, the loud talking condition identification performance is quite low. Comparing the loud talking condition identification performance with that of the shouted talking condition, it is clear that the loud talking condition identification performance is almost twice as that for the shouted talking condition. This is because the shouted talking condition consists of two components: loud talking condition and noise [10]. The percentage of confusion of the loud talking condition based on HMM1s is significant with each of the happy talking condition (16%) and the angry talking condition (15%).

Table V shows that HMM2s significantly enhance the loud talking condition identification performance compared to that using HMM1s. The improvement rate is 11.1%.

The angry talking condition identification performance based on HMM1s is very low. The identification performance for this talking condition is the second lowest identification performance in our results. The identification performance of each of the shouted talking condition and the angry talking condition is close to each other since the shouted talking condition can not be entirely separated from the angry talking condition in real life. Table II shows that the angry talking condition is highly confusable with each of the shouted and loud talking conditions.

TABLE I. TALKING CONDITION IDENTIFICATION PERFORMANCE BASED ON HMM1s

| Talking condition | Males | Females | Average |
|---|---|---|---|
| Neutral | 99% | 99% | 99% |
| Shouted | 29% | 31% | 30% |
| Loud | 52% | 56% | 54% |
| Angry | 37% | 39% | 38% |
| Happy | 58% | 60% | 59% |
| Fear | 48% | 50% | 49% |

TABLE II. CONFUSION MATRIX BASED ON HMM1s

| | Percentage of confusion of a test talking condition with the other talking conditions | | | | | |
|---|---|---|---|---|---|---|
| Model | Neutral | Shouted | Loud | Angry | Happy | Fear |
| Neutral | 99% | 2% | 1% | 3% | 1% | 4% |
| Shouted | 0% | 30% | 7% | 28% | 10% | 7% |
| Loud | 1% | 34% | 54% | 17% | 25% | 17% |
| Angry | 0% | 31% | 15% | 38% | 0% | 15% |
| Happy | 0% | 0% | 16% | 4% | 59% | 8% |
| Fear | 0% | 3% | 7% | 10% | 5% | 49% |

TABLE III. TALKING CONDITION IDENTIFICATION PERFORMANCE BASED ON HMM2s

| Talking condition | Males | Females | Average |
|---|---|---|---|
| Neutral | 99% | 99% | 99% |
| Shouted | 37% | 39% | 38% |
| Loud | 58% | 62% | 60% |
| Angry | 44% | 48% | 46% |
| Happy | 62% | 66% | 64% |
| Fear | 54% | 56% | 55% |

TABLE IV. CONFUSION MATRIX BASED ON HMM2s

| | Percentage of confusion of a test talking condition with the other talking conditions | | | | | |
|---|---|---|---|---|---|---|
| Model | Neutral | Shouted | Loud | Angry | Happy | Fear |
| Neutral | 99% | 2% | 1% | 3% | 1% | 4% |
| Shouted | 0% | 38% | 8% | 25% | 8% | 6% |
| Loud | 1% | 29% | 60% | 12% | 22% | 14% |
| Angry | 0% | 28% | 12% | 46% | 0% | 13% |
| Happy | 0% | 0% | 13% | 4% | 64% | 8% |
| Fear | 0% | 3% | 6% | 10% | 5% | 55% |

TABLE V. AVERAGE IMROVEMENT RATE OF USING HMM2s OVER HMM1s

| Model | Neutral | Shouted | Loud | Angry | Happy | Fear |
|---|---|---|---|---|---|---|
| % | 0 | 26.7 | 11.1 | 21.1 | 8.5 | 12.2 |

The angry talking condition identification performance based on HMM2s has been significantly improved. The improvement rate is 21.1%.

Our results in this work show that the happy talking condition identification performance based on HMM1s is quite low. The reason for this low talking condition identification performance is that when people are happy, they do not usually talk clearly. Consequently, there will be confusion and ambiguity in their speech production. The percentage of confusion of the happy talking condition is significant with the loud talking condition (25%) and with the shouted talking condition (10%).

The happy talking condition identification performance based on HMM2s has been remarkably improved. The improvement rate is 8.5%.

The fear talking condition identification performance based on HMM1s is worst than that for the happy talking condition. This is because when people speak under the fear mode, they produce speech with more ambiguity and less clarity than that under the happy mode. Table II shows that the fear talking condition has confusion with every talking condition especially with each of the loud and angry talking conditions.

It is evident from Table III that the fear talking condition identification performance has been drastically improved based on HMM2s compared to HMM1s. Table 5 shows that the improvement rate is 12.2%.

This is the first known investigation into HMM2s evaluated for talking condition identification systems. This work shows that HMM2s significantly enhance the recognition performance of text-dependent and speaker-dependent talking condition identification systems based on HMM2s compared to HMM1s. This is because in HMM2s, the state-transition probability at time $t$+1 depends on the states of the Markov chain at times $t$ and $t$-1. Therefore, the underlying state sequence in HMM2s is a second-order Markov chain where the stochastic process is specified by a 3-D matrix. On the other hand, in HMM1s it is assumed that the state-transition probability at time $t$+1 depends only on the state of the Markov chain at time $t$. Therefore, in HMM1s the underlying state sequence is a first-order Markov chain where the stochastic process is specified by a 2-D matrix. The stochastic process that is specified by a 3-D matrix gives more accurate talking condition recognition performance than that specified by a 2-D matrix.

VII. CONCLUDING REMARKS

As a conclusion of this work, the improvement rate of talking condition identification performance based on HMM2s is limited. The reason is that HMM2s are inefficient models to integrate observations from different modalities because observations within HMM2s have to be at a constant rate; therefore, it is impossible to look at the dynamic behavior over the course of a word if, meanwhile, we want to observe acoustic events at a much smaller time scale.